\documentclass[twocolumn,showpacs,preprintnumbers,amsmath,amssymb]{revtex4}
\usepackage{graphicx}

\begin{document}
\title{\bf Generalized Friedberg-Lee model for CP violation in neutrino physics}
\author{N. Razzaghi}
\author{S.S. Gousheh}%
\email{email: ss-gousheh@sbu.ac.ir}
\affiliation{%
Department of Physics, Shahid Beheshti University G.C., Evin, Tehran
19839, Iran
}%
\date{\today}

\begin{abstract}
We propose a phenomenological model of Dirac neutrino mass operator based on the Fridberg-Lee (FL) neutrino mass model to include CP violation. By considering the most general set of complex coefficients, and imposing the condition that the mass eigenvalues are real, we find a neutrino mass matrix which is non-hermitian, symmetric and magic. In particular, we find that the requirement of obtaining real mass eigenvalues by transferring the residual phases to the mass eigenstates self-consistently, dictates the following relationship between the imaginary part of the mass matrix elements $B$ and the parameters of the FL model: $B=\pm\sqrt{\frac{3}{4}(a-b_{r})^{2}\sin^{2}2\theta_{13}\cos^{2}\theta_{12}}$. We obtain inverted neutrino mass hierarchy, $m_{3}=0$. Making a correspondence between our model and the experimental data produces stringent conditions on the parameters as follows: $35.06^{\circ}\lesssim\theta_{12}\lesssim36.27^{\circ}$, $\theta_{23}= 45^{\circ}$, $7.27^{\circ}\lesssim\theta_{13}\lesssim11.09^{\circ}$, and $82.03^{\circ}\lesssim\delta\lesssim85.37^{\circ}$. We get mildly broken $\mu-\tau$ symmetry, which reduces the resultant neutrino mixing pattern from tribimaximal (TBM) to trimaximal (TM). The CP violation as measured by the Jarlskog parameter is restricted by $0.027\lesssim J\lesssim0.044$.
\end{abstract}
\maketitle

\section{Introduction}
In 1957 B. Pontecorvo \cite{Pontecorvo} suggested that, similar to the $k$ meson system, neutrino weak eigenstates are not mass eigenstates, but are superpositions of its mass eigenstates. Therefore as neutrinos propagate they would undergo oscillations. The full theory of neutrino oscillation was worked out in several papers \cite{oscillation}.

The results of the neutrino oscillation experiments (The solar \cite{Solar}, atmospheric \cite{atmospheric}, reactor \cite{reactor} and accelerator \cite{accelerator} neutrino experiments) have shown that mixing among three generations in the lepton sector exists, analogous to  the quark mixing, and at least two neutrinos are massive. After the discovery of neutrino oscillations, there have been many works for determining the values of the neutrino mass-squared differences and the mixing angles that relate the flavor eigenstates to the mass eigenstates. The lepton mixing matrix in the standard parametrization is given by \cite{PMNS}, \cite{para}:
\begin{eqnarray}\label{e1}
U_{PMNS}&=&\left(\begin{array}{ccc}1 & 0 & 0\\
0 & c_{23} & s_{23}\\0
& -s_{23} & c_{23}\end{array}\right)\left(\begin{array}{ccc}c_{13} & 0 & s_{13}e^{-i\delta}\\
0 & 1 & 0\\-s_{13}e^{i\delta}
& 0 & c_{13}\end{array}\right)\nonumber\\&\times&\left(\begin{array}{ccc}c_{12} & s_{12} & 0 \\
-s_{12} & c_{12} & 0\\0
& 0 & 1\end{array}\right)\left(\begin{array}{ccc}e^{i\rho} & 0 & 0 \\
0 & 1 & 0\\0
& 0 & e^{i\sigma}\end{array}\right),
\end{eqnarray}
where $c_{ij}\equiv\cos\theta_{ij}\text{ and }s_{ij}\equiv\sin\theta_{ij}$ (for $ij=12, 13 \text{ and } 23 $). The phase $ \delta $ is called the Dirac phase, analogous to the CKM phase, and the phases $\rho$ and $ \sigma$ are called the Majorana phases that are relevant for Majorana neutrinos. However, we should mention that recently the advantages of the original symmetrical form of the parameterizations of the lepton mixing matrix is discussed in \cite{para}. The results of the Daya Bay and RENO Collaborations has shown that $\theta_{13}=0$ is now rejected at a significance level higher than $8\sigma$. Analysis of current experimental data \cite{experimental} yields $31.31^{\circ}\lesssim\theta_{12}\lesssim37.46^{\circ}$, $38.64^{\circ}<\theta_{23}<53.13^{\circ}$ and $7.27^{\circ}<\theta_{13}<11.09^{\circ}$ at the $3\sigma$ confidence level. Also A combined analysis of the data coming from T2K, MINOS, Double Chooz, and Daya Bay experiments shows that the best-fit value of $\theta_{13}$ is $\sin^{2}\theta_{13}=0.026(0.027)_{-0.004}^{+0.003}$ for normal (or inverted) mass hierarchy.

One important aspect of the neutrino mixing phenomena is that it could, in principle, provide new keys for the understanding of the flavor problem, particularly since it contains large mixing angles in contrast to the quark sector. Also, the disparity between the neutrino and the charged lepton masses is more pronounced than the analogous one in the quark sector. Therefore, the mass and mixing problem in the lepton sector is a fundamental problem. Also some interesting questions to be solved by future experiments are what are the masses of neutrinos, how close to $45^{\circ}$ is the 2-3 mixing angle, what are the values of three CP-violating phases of the PMNS matrix (i.e., the Dirac phase $ \delta $ and the Majorana phases $ \rho \text{ and } \sigma $).
The simplest way to introduce massive neutrinos is to add right-handed chiral neutrino fields to the standard model and to introduce the neutrino masses in the same way as for the quarks and charged leptons.
In this paper, we mainly focus on the Dirac neutrinos. However, we should mention that the determination of the nature of neutrinos is still a controversial subject which could eventually be decided by experimental observation, such as non-zero magnetic dipole moment of neutrinos ruling out majorana neutrinos or neutrino-less double beta decay ruling out Dirac neutrinos. We believe that at this point in time the study of both types of neutrinos is justified (see for example \cite{Farzan}). A Dirac mass term for the neutrinos and charged leptons are written as
 \begin{equation}\label{e2}\vspace{.2cm}
  {\cal L_\mathbf{m}}= {-\bar{\ell}_{{Li}}}{\cal
M_\mathbf{e}}^{ij}\ell_{Rj}-{\bar{\nu}_{{Li}}}{\cal
M_\mathbf{D}}^{ij}\nu_{Rj}+h.c.,
\end{equation}
\par A successful phenomenological neutrino mass model with the flavor symmetry that is suitable for the Dirac neutrinos was proposed by Friedberge and Lee (FL) \cite{FL}. In this model the mass eigenstates of three charged leptons are identified with their flavor eigenstates. Therefore neutrino mixing matrix can be simply described by a $ 3\times3 $ unitary matrix U which transforms the neutrino mass eigenstates to the flavor eigenstates, $ (\nu_{e},\nu_{\mu},\nu_{\tau}).$ As we shall show in the pure FL model, one of the neutrino masses is exactly zero, partially justifying the smallness of neutrino masses. Morover when $\mu-\nu$ symmetry is assumed, the matrix $U$ reduces to the experimentally favored $U_{TBM}$.

The Dirac neutrino mass operator of the FL model can be written as
\begin{eqnarray}\label{e3}\vspace{.5cm}
  {\cal M_\mathbf{FL}}&=&a\left(\bar{\nu}_{\tau}-\bar{\nu}_{\mu}\right)\left(\nu_{\tau}-\nu_{\mu}\right)
+ b\left(\bar{\nu}_{\mu}-\bar{\nu}_{e}\right)\left(\nu_{\mu}-\nu_{e}\right)\nonumber\\
&+& c\left(\bar{\nu}_{e}-\bar{\nu}_{\tau}\right)\left(\nu_{e}-\nu_{\tau}\right)\nonumber\\
&+& m_{0}\left(\bar{\nu}_{e}\nu_{e}+\bar{\nu}_{\mu}\nu_{\mu}+\bar{\nu}_{\tau}\nu_{\tau}\right).
\end{eqnarray}
All the parameters in this model ($a,b,c$ and $m_{0}$) are assume to be real. For $m_{0}=0$, this Lagrangian has the following symmetry $
 \nu_{e}\rightarrow\nu_{e}+z $, $\nu_{\mu}\rightarrow\nu_{\mu}+z $, and $  \nu_{\tau}\rightarrow\nu_{\tau}+z $, where $z$ is an element of the Grassman algebra. For constant $z$, this symmetry is called FL symmetry \cite{FL} in which case the kinetic term is also invariant. However the other terms of the electroweak Lagrangian do not have such symmetry. The $ m_{0} $ term  breaks this symmetry explicitly. However we may add that the FL symmetry leads to a magic matrix and this property is not spoiled by the $m_{0}$ term. The magic property has many manifestations which we shall discus in details. Also it has been reasoned that the FL symmetry is the residual symmetry of the neutrino mass matrix after the $SO(3)\times U(1)$ flavor symmetry breaking \cite{FL2}. The mass matrix can be displayed by,
 \begin{equation}\label{e4}
M_{FL} =\left(\begin{array}{ccc}b+c+m_{0} & -b & -c\\
-b & a+b+m_{0} & -a\\-c & -a & a+c+m_{0}\end{array}\right),
\end{equation}
where $a \propto\left(Y_{\mu\tau}+Y_{\tau\mu }\right)$, $b \propto\left(Y_{e\tau}+Y_{\tau e }\right)$ and $  \ c \propto\left(Y_{\tau e}+Y_{e\tau }\right) $ and $Y_{\alpha\beta}$ denote the Yukawa coupling constant. The proportionality constant is the expectation value of the Higgs field. It is apparent that $ M_{FL} $ possesses exact $\mu-\tau$ symmetry only when $b=c$. Setting $b=c$ and using the hermiticity of $M_{FL}$, a straight forward diagonalization procedure yields $U^{T}_{TBM}M_{FL}U_{TBM}=\text{ Diag }\{m_{1},m_{2},m_{3}\} $ where,
\begin{eqnarray}\label{e6}\vspace{.2cm}
m_{1}&=&3b+m_{0}\nonumber\\m_{2}&=&m_{0}\nonumber
\\m_{3}&=&2a+b+m_{0},
\end{eqnarray}
and the experimentally favored  tribimaximal (TBM) neutrino mixing matrix can be reproduced and is given by
\begin{equation}\label{e5}
U_{TBM} =\left(\begin{array}{ccc}\frac{2}{\sqrt{6}} & \frac{1}{\sqrt{3}} & 0\\
-\frac{1}{\sqrt{6}} & \frac{1}{\sqrt{3}} & \frac{1}{\sqrt{2}}\\-\frac{1}{\sqrt{6}} & \frac{1}{\sqrt{3}} & -\frac{1}{\sqrt{2}}\end{array}\right).
\end{equation}
Obviously the requirement that all of the mass eigenvalues are positive puts conditions on the parameters of this model. In particular  $m_{\mathbf{0}}$ must be positive. It is interesting to note that $U_{TBM}$ was first proposed on theoretical grounds by Harrison, Perkins and Scott in 2002 \cite{Harrison}.
For a general exact TBM neutrino mixing, regardless of the model, the mixing angles are $\theta_{12}\approx35.3^{\circ}$, $\theta_{23}=45^{\circ}$, $\theta_{13}=0$ and the CP-violating phases can be chosen to be zero.
In general in order to have CP-violation, the necessary condition is $\delta\neq0$ and $\theta_{13}\neq0$. In this model these conditions necessarily mandate that $\mu-\tau$ symmetry should be broken. Another interesting question is whether $\theta_{23}=45^{\circ}$ remains correct after the $\mu-\tau$ symmetry breaking.

There are four independent CP-even quadratic invariants, which can conveniently be chosen as $U^{\ast}_{11}U_{11}, U^{\ast}_{13}U_{13}, U^{\ast}_{21}U_{21}$ and $ U^{\ast}_{23}U_{23} $ and three independent CP-odd quadratic invariants \cite{quadratic},
\begin{eqnarray}\label{e7}\vspace{.2cm}
J&=&\Im(U_{11}U^{\ast}_{12}U^{\ast}_{21}U_{22})\nonumber\\ I_{1}&=&\Im[(U^{\ast}_{11}U_{12})^{2}]\nonumber\\I_{2}&=&\Im[(U^{\ast}_{11}U_{13})^{2}].
\end{eqnarray}
The Jarlskog rephasing invariants $J$ \cite{J}, is relevant for CP violation in lepton number conserving processes like neutrino oscillations. $I_{1}$ and $I_{2}$ are relevant for CP violation in lepton number violating processes like neutrinoless double beta decay. Oscillation experiments cannot distinguish between the Dirac and Majorana neutrinos. The detection of neutrinoless double beta decay would provide direct evidence of lepton number non-conservation and the Majorana nature of neutrinos. Many theoretical and phenomenological works have discussed  massive neutrino models that break $\mu-\tau$ symmetry as a prelude to CP violation\cite{theoretical}.

In this paper we generalize the FL model by introducing complex parameters which can ultimately be linked to complex Yukawa coupling constants. We concentrate on the massive FL Dirac model, imposing the obvious constraint that mass eigenvalues be real. Using this model we obtain CP violation, mild  $\mu-\tau$ symmetry breaking  \cite{xing}, and inverted mass hierarchy for neutrinos.  Moreover the measures of these two symmetry breakings turn out to be proportional to each other. This paper is organized as follows. In section 2, We introduce our model and show how the constraint of reality of masses along with the minimal breaking of $\mu-\tau$ symmetry, and the overall self consistency of the model produces relationships between the free parameters of the model. We find that in our model $0<\theta_{13}<24^{\circ}$, $35.24^{\circ}<\theta_{12}<39.20^{\circ}$, $\theta_{23}=45^{\circ} $ and $71.56^{\circ}<\delta<\frac{\pi}{2}$. Notice that we have ruled out the case $\delta=\frac{\pi}{2}$ as we shall explain. %In section 3, we discuss the nature of the Yukawa coupling constants on more general grounds.
In section 3, we show that our model is in general consistent with the experimental data, and show that implementing all of the constraints coming from the experimental data, severely restricts the parameters of our model, an in fact almost pinpoints the relevant parameters. Section 4 is devoted to a summary.

\section{Model }
In this section, we generalize the FL model by adding complex Yukawa coupling constants, in order to obtain CP violation. This is accomplished by obtaining non-zero values for $ \sin\theta_{13} $ and $\delta$. We first let all of the coefficients in the $M_{FL}$ matrix Eq.\,(\ref{e4}) except $m_{0}$ be complex. However we demand the eigenvalues of the mass matrix to be real. We find that only one particular choice allows for minimal breaking of $\mu-\tau$ symmetry, {\em i.e.} $(a\in\Re;~ b, c\in\mathbb{C}$ and $b=c^{\star})$. This requirement leads to a non-hermitian mass matrix.  For simplicity of notation we define the parameters as follows: $\Re\left(b\right)=\Re\left(c\right)= b_{r} \text{ and }\Im\left(b\right)=-\Im\left(c\right)= B$. The parameters indicating the measure of CP violation and $\mu-\tau$ symmetry breaking turn out to be proportional to $B$ and therefore we expect it to be small.

The neutrino mass matrix $ M^{\prime}_{\nu}$ is given by
\begin{eqnarray}\label{e8}
M^{\prime}_{\nu}&=&\left(\begin{array}{ccc}2b_{r}+m_{0} & -b_{r} & -b_{r}\\
-b_{r} & a+b_{r}+m_{0} & -a\\-b_{r}
& -a & a+b_{r}+m_{0}\end{array}\right)\nonumber\\&+&iB\left(\begin{array}{ccc}0 & -1 & 1\\
-1 & 1 & 0\\1 & 0 & -1\end{array}\right),
\end{eqnarray}

Notice that $M^{\prime}_{\nu}$ and $M_{FL}$ are both magic and symmetric matrices since they both commute with the magic $S$ matrix defined bellow,
\begin{equation}\label{e9}
S =\left(\begin{array}{ccc}F & T & T\\
T & F & T\\T & T & F\end{array}\right).
\end{equation}
Therefore one of the eigenstates is $(\frac{1}{\sqrt{3}} , \frac{1}{\sqrt{3}} , \frac{1}{\sqrt{3}})$, and we choose it to be $\nu_{2}$ in order to be consistent with Eq.\,(\ref{e5}). In the special case $F=T=\frac{1}{3}$, $S=D$ where $D$ is called the Democracy Operator.
It is obvious that the real part of $M^{\prime}_{\nu}$ has CP and $\mu-\tau$ symmetry, separately. The imaginary part of $M^{\prime}_{\nu}$ breaks both of these symmetries, while preserving the product of these two operations. Therefore $ {|M^{\prime}_{\nu}}|^{2}=M^{\prime}_{\nu}{M^{\prime}_{\nu}}^{\dag}$ is invariant under simultaneous CP and $\mu-\tau$ reflection operations, which is called the mutativity operation \cite{219}. This reduction of the symmetry, causes the symmetry of neutrino mixing matrix to reduce from TBM to TM.

A naive diagonalization of $M^{\prime}_{\nu}$ yields,
\begin{eqnarray} \vspace{.2cm}\label{e11}
\breve{m}_{1}&=&(a+2b_{r}+m_{0})+\sqrt{(a-b_{r})^{2}-3B^{2}}\nonumber\\\breve{m}_{2}&=&m_{0},\nonumber\\\breve{m}_{3}&=&(a+2b_{r}+m_{0})-\sqrt{(a-b_{r})^{2}-3B^{2}}.
\end{eqnarray}
However the usual diagonalization procedure is correct only for hermitian matrices, where a similarity transformation by a unitary operator, i.e. $M^{\prime}_{diag}=U^{\dag}M^{\prime}_{\nu}U$,  diagonalizes the matrix. Therefore the results indicated in Eq.\,(\ref{e11}) are correct only in the limit $B\rightarrow0$, and the results in this limit suffice for our analysis to follow. Comparing our results in this limit with those shown in Eq.\,(\ref{e6}), we conclude $ a<b_{r} $.

Since $M^{\prime}_{\nu}$ is a non-hermitian matrix, we need two distinct unitary matrices $U$ and $V$ to diagonalize it. These matrices can be easily obtained by diagonalizing $ M^{\prime}_{\nu}{M^{\prime}_{\nu}}^{\dag} $ and $ {M^{\prime}_{\nu}}^{\dag}M^{\prime}_{\nu} $, separately. $U$ and $V$ are the conventional transformation matrices for the left-handed and right-handed neutrinos, respectively. we do not display the explicit form of $U$ and $V$ and only mention that $V = U^{\ast}$. The resulting correct diagonal matrix is obtained by $M^{\prime}_{diag}=U^{\dag}M^{\prime}_{\nu}V$ and its elements are as follows,
\begin{widetext}
\begin{eqnarray} \vspace{.2cm}\label{e111}
m'_{1}&=&\frac{iB(a-b_{r})+3B^{2}+(a+2b_{r}+m_{0})^{2}-(a-b_{r}+iB)\sqrt{3B^{2}+(a+2b_{r}+m_{0})^{2}}}{a+2(b_{r}+iB)+m_{0}}\nonumber\\m'_{2}&=&m_{0},\nonumber\\m'_{3}&=&\frac{iB(a-b_{r})+3B^{2}+(a+2b_{r}+m_{0})^{2}+(a-b_{r}+iB)\sqrt{3B^{2}+(a+2b_{r}+m_{0})^{2}}}{a+2(b_{r}+iB)+m_{0}}.
\end{eqnarray}
\end{widetext}
After the diagonalization we find that only $m'_{1}$ and $m'_{3}$ are complex. We can extract the phases and transfer them to the mass eigenstates in the Dirac case \cite{phase}. The most general form the diagonal mass matrix can be written as,
\begin{equation}\label{ereal}
M^{\prime}_{diag}=e^{i\alpha}e^{i\beta\lambda_{3}}e^{i\gamma\lambda_{8}}M'^{real}_{diag}.
\end{equation}
In our model $\alpha$ automatically turns out to be zero. We would dispense with the overall phase even if it was not zero. Using the fact that $m'_{2}$ is real, we obtain $\beta=\gamma$. This implies that the $\arg(m'_{1})=-\arg(m'_{3})=2\beta$. Using these conditions in Eq.(11) we obtain
\begin{equation}\label{e12}
B=\pm\sqrt{\frac{-\left(2a+b_{r}+m_{0}\right)\left(3b_{r}+m_{0}\right)}{3}}
\end{equation}
Note that disregarding the overall phase amounts to the following: $Det(M^{\prime}_{diag})$ is real and $Det (U)=1$\cite{phase}.

\begin{center}
\begin{figure}[th] \includegraphics[width=8cm]{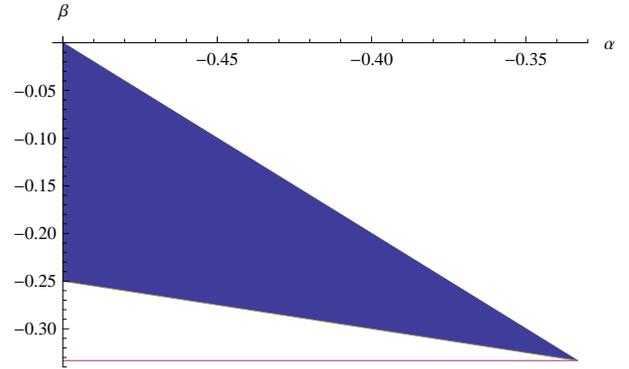}\caption{\label{fig.1} \small
   CP violation is possible only in the right-angled triangle in our parameter space. The axis are defined by $\alpha\equiv\frac{a}{m_{0}}$ and $\beta\equiv\frac{b_{r}}{m_{0}}$. The dark triangle displays the allowed region within our model. (The line above the base of the triangle is given by $2b_{r}+a+m_{0}=0$)}
  \label{geometry}
\end{figure}
\end{center}

From the requirement of the reality of B we obtain $ -\frac{m_{0}}{3}\leq b_{r} \leq -(2a+m_{0}) $. Notice that the lower bound of $b_{r}$ is simply a check on the condition  $m_{1}>0$ in Eq.\,(\ref{e6}). The requirement that in the limit $B\rightarrow0$, $U$ and $V$ should both approach $U_{TBM}$ given in Eq.\,(\ref{e5}), yields $2b_{r}+a+m_{0}>0$. Combining the condition for reality of $B$ with $ a<b_{r} $, we obtain $3a+m_{0}<0$. From this and the overall symmetry of the F.L model we conclude that $b_{r}<0$. This conclusion is consistent with  the results of experiments on solar neutrino oscillation which indicate that $m_{2}>m_{1}$. Notice that the occurrence of CP violation is possible only in a restricted region in the $a$-$b_{r}$ plane where $B\neq0$. Figure (\ref{fig.1}) illustrates the region of the parameter space where CP violation occurs.

Substituting the expression for $B$ given by Eq.\,(\ref{e12}) into the expression we have obtained for $U$ by diagonalizing $ M^{\prime}_{\nu}{M^{\prime}_{\nu}}^{\dag} $, we obtain
\begin{widetext}
\begin{eqnarray} \vspace{.2cm}\label{e13}
U_{11}&=&\sqrt{\frac{1}{\left(a-b_{r}\right)\left(3a+m_{0}\right)}}\left(\frac{-\left(3b_{r}+m_{0}\right)}{\sqrt{6}}
-i\sqrt{\frac{-\left(2a+b_{r}+m_{0}\right)\left(3b_{r}+m_{0}\right)}{2}}\right),
\nonumber\\U_{12}&=&U_{22}=U_{32}=\frac{1}{\sqrt{3}},
\nonumber\\U_{13}&=&\sqrt{\frac{-1}{(a-b_{r})(-a+4b_{r}+m_{0})}}\left(\frac{(2a+b_{r}+m_{0})}{\sqrt{6}}
+i\sqrt{\frac{-(2a+b_{r}+m_{0})(3b_{r}+m_{0})}{2}}\right),
\nonumber\\U_{21}&=&\sqrt{\frac{1}{\left(a-b_{r}\right)\left(3a+m_{0}\right)}}\left(\frac{3(a+b_{r})+2m_{0}}{\sqrt{6}}
+i\sqrt{\frac{-(2a+b_{r}+m_{0})(3b_{r}+m_{0})}{2}}\right),
\nonumber\\U_{23}&=&\sqrt{\frac{-1}{(a-b_{r})(-a+4b_{r}+m_{0})}}\left(\frac{-(a+5b_{r}+2m_{0})}{\sqrt{6}}
-i\sqrt{\frac{-(2a+b_{r}+m_{0})(3b_{r}+m_{0})}{2}}\right),
\nonumber\\U_{31}&=&\sqrt{\frac{1}{\left(a-b_{r}\right)\left(3a+m_{0}\right)}}\left(\frac{-(3a+m_{0})}{\sqrt{6}}\right),
\nonumber\\U_{33}&=&\sqrt{\frac{-1}{(a-b_{r})(-a+4b_{r}+m_{0})}}\left(\frac{-a+4b_{r}+m_{0}}{\sqrt{6}}\right).
\end{eqnarray}
\end{widetext}
Notice that our generalized $M^{\prime}_{\nu}$ given in Eq.\,(\ref{e8}), has retained its magic property, since we have insisted on having real mass eigenvalues. Therefore the mixing matrices $U$ and $V$ are necessarily TM, which requires for example $|\nu_{2}\rangle=\frac{1}{\sqrt{3}}(\nu_{e}+\nu_{\mu}+\nu_{\tau})$ and their first and third columns add up to zero \cite{sanjeev}. However the TBM structure is broken down to TM since $U_{13}\neq0$, and therefore the exact $\mu-\tau$ symmetry is broken. As we shall show this symmetry is only softly broken. We can conclude from the TM nature of the $U$ matrix that $3(a+b_{r})+m_{0}>0$.

Comparing Eq.\,(\ref{e13}) with Eq.\,(\ref{e1}), we immediately obtain all of the mixing angles ($ \theta_{13} , \theta_{12}, \theta_{23} $) and the CP-violating phase in terms of $a, b_{r}$ and $m_{0}$ as follows
\begin{eqnarray}\label{e14}
\sin^{2}\theta_{13}&=&\frac{2a+b_{r}+m_{0}}{3(a-b_{r})},
\nonumber\\\sin^{2}\theta_{12}&=&\frac{1}{3\cos^{2}\theta_{13}}=\frac{a-b_{r}}{a-4b_{r}-m_{0}},
\nonumber\\\sin^{2}\theta_{23}&=&\frac{1}{2},
\nonumber\\\tan\delta&=&\sqrt{\frac{-3(3b_{r}+m_{0})}{2a+b_{r}+m_{0}}}.
\end{eqnarray}
In other words, the phase difference between $b$ and $c$ results in a kind of $\mu-\tau$ symmetry breaking with a manifestation $\theta_{13}\neq0$, while maintaining $\theta_{23}=\frac{\pi}{4}$ ($|U_{23}|=|U_{33}|$).
From Eq.\,(\ref{e1}), Eq.\,(\ref{e12}) and Eq.\,(\ref{e14}) we obtain,
\begin{equation}\label{e15}
B=\pm\sqrt{\frac{3}{4}(a-b_{r})^{2}\sin^{2}2\theta_{13}\cos^{2}\theta_{12}}.
\end{equation}

Using  Eq.\,(\ref{e13}) and the transformation $ U^{\dag} M^{\prime}_{\nu} V $, or by substituting Eq.\,(\ref{e12}) in to Eq.\,(\ref{e111}), it is easy to obtain the three neutrino masses
\begin{equation}\label{e16}
M^{\prime}_{\text{diag }} =e^{i\beta\lambda_{3}}e^{i\beta\lambda_{8}}\left(\begin{array}{ccc}-2(a-b_{r}) & 0 & 0 \\
0 & m_{0} & 0\\ 0 & 0 & 0\end{array}\right)
\end{equation}
 where,
\begin{equation} \vspace{.2cm}\label{e17}
\beta=\arctan\left(-\frac{3B}{3(a+b_{r})+2m_{0}}\right),
\end{equation}
Notice that we get inverse hierarchy for neutrino masses. In the Dirac case one can choose, without loss of generality, the phases of the mass  eigenstates ( $\acute{\nu}^{i}_{L,R}$ ) so that $M^{\prime}_{\text{diag }}$ reduces to $M'^{real}_{diag}$ as in Eq.\,(\ref{ereal}).

Since $M^{\prime}_{\nu}$ is a symmetric matrix, it could also be used as a Majorana mass matrix. If we work with Majorana neutrinos, every element in the mixing matrix will be same as the Dirac case. The phases can be rewritten as,
\begin{equation} \vspace{.2cm}\label{emajrana}
e^{i\beta\lambda_{3}}e^{i\beta\lambda_{8}}= \left(\begin{array}{ccc}e^{i2\beta} & 0 & 0 \\
0 & 1 & 0\\0
& 0 & e^{-i2\beta}\end{array}\right).
\end{equation}
In the Majorana case only the phase factor $e^{-2i\beta}$ in Eq.\,(\ref{emajrana}) can be rotated away into the charged lepton sector and the other phase factor remains. These phases can be transferred to $U_{PMNS}$, and comparing Eq.\,(\ref{e1}) and Eq.\,(\ref{emajrana}) one can conclude that the $\rho=-\sigma=-\beta$. These phases also contribute to the CP violation. Therefore in Majorana case we obtain three non-zero CP violating phases $\delta$ and $\rho=-\sigma$, with inverted hierarchy, and the  masses are
\begin{eqnarray} \vspace{.2cm}\label{e19}
(m^{\prime}_{1})_{M}&=&-2(a-b_{r}),\nonumber\\(m^{\prime}_{2})_{M}&=&m_{0},\nonumber\\(m^{\prime}_{3})_{M}&=&0.
\end{eqnarray}
Therefore $\det M^{\prime}_{\nu}=0$ and $ M^{\prime}_{\nu}$ has no inverse. This shows that the use of our model for Majorana neutrinos cannot be consistent with the type I seesaw mechanism proposed in 1980 \cite{seesaw}.

A rephasing-invariant measure of CP violation in neutrino oscillation is the universal parameter $J$ \cite{J} given in Eq.(\ref{e7}), and it has a form which is independent of the choice of the Dirac or  Majorana neutrinos. Using Eq.(\ref{e13}) the expression for J simplifies to,
\begin{equation}\label{e20}
J =-\frac{B}{6(a-b_{r})}.
\end{equation}
From this expression we can conclude that the maximal CP non-conservation is not just a question of a relative phase assuming the value $\pm\frac{\pi}{2}$; the magnitudes of the coupling constants are also essential. In fact the maximal CP non-conservation does not correspond to $\sin\delta=1$ \cite{J}.
One can see that the soft breaking of $\mu-\tau$ symmetry leads to both $\theta_{13}\neq0$ and $J \neq0$, but it does not affect the favorable result $\theta_{23}=45^{\circ}$, originally resulting from the TBM mixing angles.

\section{  Comparison with experimental data }

In this section we compare the experimental data with the results obtained from our model. We do this by mapping all of the constraints obtained from the experimental data onto our parameter space, as shown in Figure (\ref{fig.2}). Note that there is a significant general overlap between the experimental data and the CP-violating part of our parameter space (as originally shown in Figure (\ref{fig.1})).

\begin{widetext}
\begin{center}
\begin{table}[h]
\centering
\begin{tabular}{|c|c|c|c|}
\hline
Parameter &  The experimental data     & The best fit ($\pm1\sigma$)   & Combining our model with the experimental data \\
\hline
$\Delta m_{21}^{2}$ & $(7.12-8.20)10^{-5}eV^{2}$ & $7.62\pm0.19$ & $m_{1}\approx(4.14-5.00)10^{-2}eV $  \\
& &  & $ m_{2}\approx(4.70-5.25)10^{-2}eV$  \\
\hline
$\Delta m_{31}^{2}$& $-(2.15-2.68)10^{-3}eV^{2}$ & $-(2.40_{-0.07}^{+0.10})$ & $m_{1}\approx(4.14-5.00)10^{-2}eV $\\
& & & $ m_{3}=0$ \\
\hline
$\sin^{2}\theta_{12}$  & $0.27-0.37$ & $0.320_{-0.017}^{+0.015}$ & $ 0.33-0.35$ \\
\hline
$\sin^{2}\theta_{13}$& $0.016-0.037$ & $0.027_{-0.004}^{+0.003}$ & $ 0.016-0.037$\\
\hline
$\sin^{2}\theta_{23}$& $0.39-0.64$ & $0.53_{-0.07}^{+0.05}$ & $ 0.5$\\
\hline
$\delta$& $0-2\pi$ & $0-2\pi$ & $82.03^{\circ}-85.37^{\circ}$\\
\hline
$J$&  &  & $0.027-0.044$\\
\hline
\end{tabular}
\caption{The allowed ranges for all parameter obtained from the experimental data and our model}\label{h}
\label{tab:PPer}
\end{table}
\end{center}
\end{widetext}
However the most restricting experimental data comes from the values of $\sin^{2}\theta_{13}$ and $m'_{1}=\sqrt{|\Delta m_{31}^{2}|}$ and $ m'_{2}=\sqrt{|\Delta m_{21}^{2}-\Delta m_{31}^{2}|}$. These restrictions are shown in the Figure (\ref{fig.2}) by shaded regions. Therefore the overlap of all of the experimental data and our model is reduced to a tiny region close to the top corner of the triangle.  In Figure (\ref{fig.3}) we have plotted the values of $|\Delta m_{31}^{2}|$ against $\sin^{2}\theta_{13}$ to elucidate this important overlap region. In TABLE \ref{h} we state all of the relevant experimental results presented at the $3\sigma$ C.L. \cite{experimental}, along with the restrictions that they impose on the parameters of our model. The values stated as the results of our model (combined with experimental data) in the last column of the table are obtained from the tiny region mentioned above which results in the following restricted values for the parameters of our model, and using $m'_{3}=0$:
\begin{eqnarray}\label{e21}
m_{0}&\approx&(4.70-5.25)10^{-2}eV ,\nonumber\\
a&\approx&-(2.3-2.6)10^{-2} eV,\nonumber\\
b_{r}&\approx&-(0.1-0.3)10^{-2}eV,\nonumber\\
B&\approx&(0.34-0.62)10^{-2}eV.
\end{eqnarray}

\begin{widetext}
\begin{center}
\begin{figure}[th] \includegraphics[width=15cm]{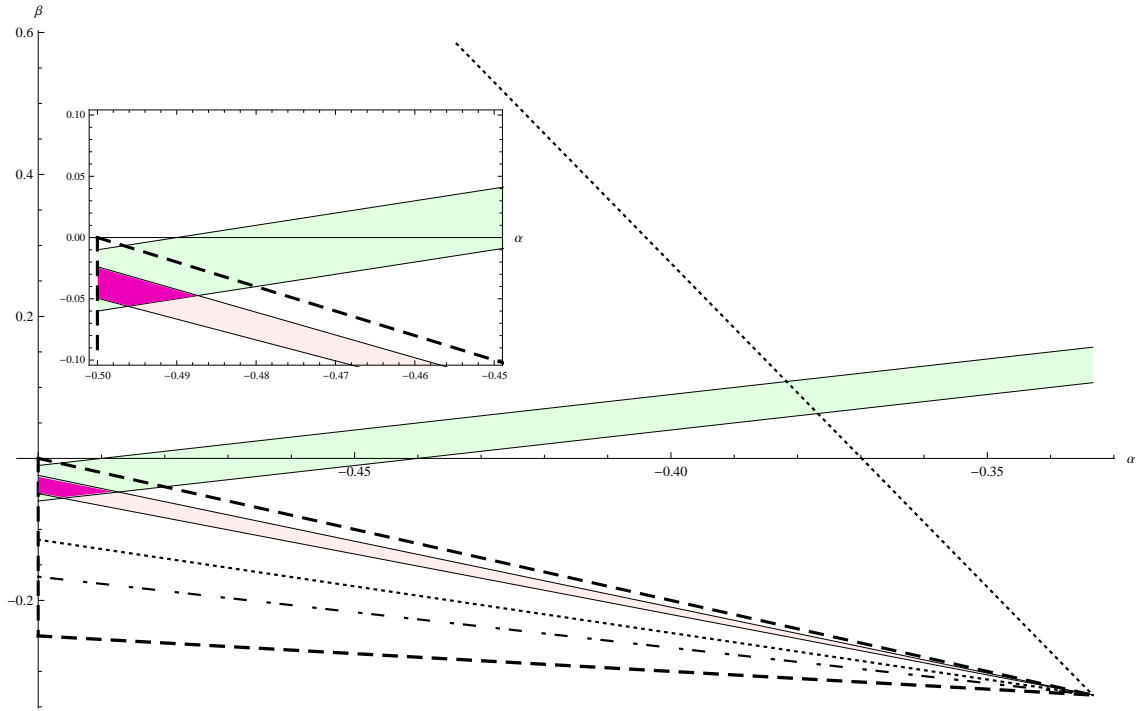}\caption{\label{fig.2} \small
   In this figure we show the theoretical region of our model for CP violation along with all available experimental data. The axes are defined by $\alpha\equiv\frac{a}{m_{0}}$ and $\beta\equiv\frac{b_{r}}{m_{0}}$. The triangle delimited by the dashed lines is identical with the colored triangle shown in Figure (\ref{fig.1}). The line immediately above the base of the triangle is the line given by $3(a+b_{r})+2m_{0}$ and in our model only the part of the mentioned triangle above this line is allowed. This line is interesting since on this line the following parameters are all constants: $J=0.083$, $ \theta_{13}=24.09^{\circ}$, $\delta=71.56^{\circ}$, while the derivative of $B$ in the direction perpendicular to this line is zero. Moving away from this line in the upward  direction, $B$ and $\theta_{13}$ decrease monotonically to zero, while $\delta$ increases to $\frac{\pi}{2}$ when the upper border of the triangle ($2a+b_{r}+m_{0}=0$) is reached. The colored region with negative slope indicates the experimentally allowed region for the $\sin^{2}\theta_{13}$. The region delimited by dotted lines indicates the experimentally allowed region for the $\sin^{2}\theta_{12}$.  The colored region bounded by two closely spaced parallel lines with positive slope are the result of the restriction coming from the experimental values for $\Delta m_{31}^{2}$ and $\Delta m_{21}^{2}$. The overlap region of all of the experimental data is indicated by the darker region and this is completely contained within the region for our model.}
  \label{geometry}
\end{figure}
\end{center}
\end{widetext}
\begin{center}
\begin{figure}[th] \includegraphics[width=9cm]{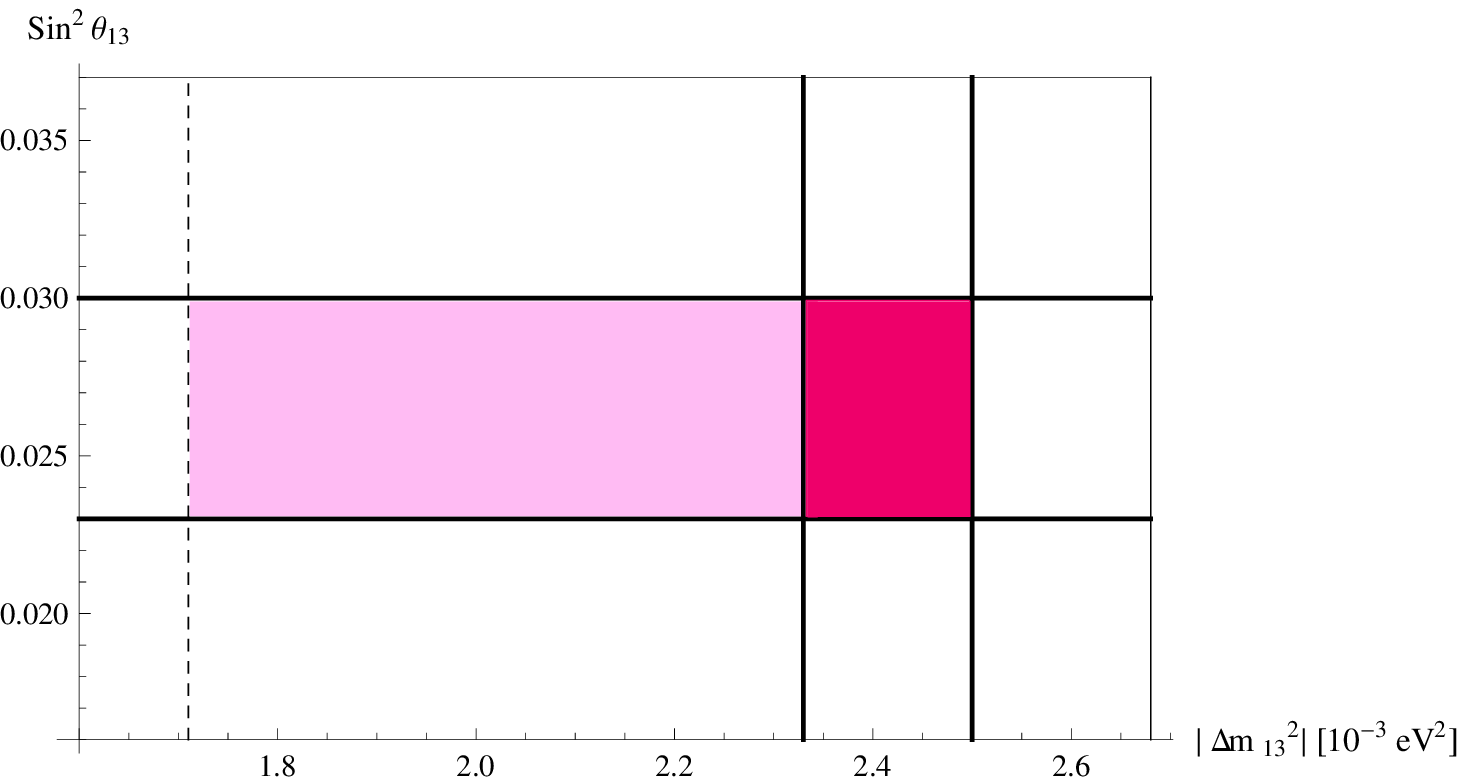}\caption{\label{fig.3} \small
   In this figure the whole experimentally allowed region of the $\sin^{2}\theta_{13}-|\Delta m_{31}^{2}|$ plane is indicated by the region bounded by the largest rectangle, which includes the axes. This whole region corresponds to the two colored regions shown in Figure (\ref{fig.2}). The best experimental fit is shown by the darker shaded region in the middle. Our model is represented by the lighter shaded region which overlaps the best experiments fit.}
  \label{geometry}
\end{figure}
\end{center}
In Figure (\ref{fig.2}) we have a special point, $ a=b_{r}=-\frac{m_{0}}{3} $ at which $B=0$. Therefore in our model we cannot have CP violation at this point, where the neutrino mass matrix reduces to the following special form
\begin{eqnarray}\label{e22}
M^{\prime}_{\nu}&=&(3b_{r}+m_{0})\left(\begin{array}{ccc}1 & 0 & 0\\
0 & 1 & 0\\0 & 0 & 1\end{array}\right)-b_{r}\left(\begin{array}{ccc}1 & 1 & 1\\
1 & 1 & 1\\1 & 1 & 1\end{array}\right)\nonumber\\&=&(3b_{r}+m_{0})\textbf{1}-b_{r}\tilde{D}.
\end{eqnarray}
the first term is identically zero and $M^{\prime}_{\nu}$ reduces to the democratic matrix $-b_{r}\tilde{D}$. $M^{\prime}_{\nu}$ can be diagonalized by the  $U_{TBM}$ matrix defined by Eq.\,(\ref{e5}), yielding
\begin{eqnarray}\label{e23}
M^{\prime}_{\text{diag}}&=&U^{T}_{TBM}[(3b_{r}+m_{0})\textbf{1}-b_{r}\tilde{D}]U_{TBM}\nonumber\\&=&\left(\begin{array}{ccc}3b_{r}+m_{0} & 0 & 0\\
0 & m_{0} & 0\\0 & 0 &3b_{r}+m_{0}\end{array}\right)
\end{eqnarray}
Note that in this case the only non-zero element is $(M^{\prime}_{\text{diag}})_{22}=m_{0}$. This matrix has attracted some attention and several authors have considered generalizations of this matrix to break the degeneracy \cite{degeneracy}.

\section{Conclusion}
In this paper we have proposed a generalization of Friedberg-Lee neutrino mass model, in which CP violation is possible. In our model the coefficients are allowed to be complex, with the constraint that the mass eigenvalues be real. We find and display the region in our parameter space where CP violation is possible. Since the parameters of our model are related to the Yukawa coupling constants, this region determines a corresponding CP violating region in the space of the Yukawa coupling constants. In this region the resulting mass matrix turns out to be non-hermitian, symmetric and magic. We find that the symmetry of the neutrino mixing matrix is reduced from TBM to TM with the implication that the $\mu-\tau$ symmetry is mildly broken. Comparing the results of our model with experimental data, we find that overlap region is very restricted and this narrows the allowed ranges for the parameters, as shown in TABLE \ref{h}. In particular we find Jarlskog parameter is restricted to $0.027\lesssim J\lesssim0.044$, which could be tested in the future experiments such as the upcoming long baseline neutrino oscillation ones). Also $35.06^{\circ}\lesssim\theta_{12}\lesssim36.27^{\circ}$, $7.27^{\circ}\lesssim\theta_{13}\lesssim11.09^{\circ}$, $\theta_{23}=45^{\circ}$ and $82.03^{\circ}\lesssim\delta\lesssim85.37^{\circ}$. We obtain the allowed ranges for the values of three masses $m_{1}\approx(4.14-5.00)10^{-2}eV $ , $m_{2}\approx(4.70-5.25)10^{-2}eV$ and $m_{3}=0$, therefore we have inverted hierarchy.

This generalization could also be used for massive Majorana neutrinos, because the generalized mass matrix is still symmetric. In Majorana case all of the parameters are identical to the Dirac case, except that there are two extra CP violation phases  $27.92^{\circ}\leq\rho=-\sigma\leq45.56^{\circ}$. However, since $\det M^{\prime}_{\nu}=0$, the mass matrix $ M^{\prime}_{\nu}$ is not invertible. This shows that the use of our model for Majorana neutrinos cannot be consistent with the type I seesaw mechanism.
\begin{acknowledgments} \label{Calculation}
We would like to thank the research office of the Shahid Beheshti University for financial support.
\end{acknowledgments}

 \end{document}